# Bell's theorem and the issue of determinism and indeterminism

Michael Esfeld
University of Lausanne, Department of Philosophy
Michael-Andreas.Esfeld@unil.ch

**Abstract**

The paper considers the claim that quantum theories with a deterministic dynamics of objects in ordinary space-time, such as Bohmian mechanics, contradict the assumption that the measurement settings can be freely chosen in the EPR experiment. That assumption is one of the premises of Bell's theorem. I first argue that only a premise to the effect that what determines the choice of the measurement settings is independent of what determines the past state of the measured system is needed for the derivation of Bell's theorem. Determinism as such does not undermine that independence (unless there are particular initial conditions of the universe that would amount to conspiracy). Only entanglement could do so. However, generic entanglement without collapse on the level of the universal wave-function can go together with effective wave-functions for subsystems of the universe, as in Bohmian mechanics. The paper argues that such effective wave-functions are sufficient for the mentioned independence premise to hold.

## 1. Introduction

As is by now well-established in the literature, the derivation of Bell's theorem is based on two premises: a principle of locality and a principle of independence according to which the choice of the variables to be measured on the system in question is independent of what influences the past state of the system. Whereas the locality premise is discussed in all the literature on quantum physics, the independence premise has received less attention, although it is also significant: it is alleged that this premise implies some sort of indeterminism in nature. Most recently, for instance, the theorem of Colbeck and Renner (2011) takes such an implication for granted.

The purpose of the present paper is to assess this alleged implication. The interest of such an investigation is obvious. In case this implication stands firm, then if one has reasons to maintain that nature operates in a deterministic manner, the commitment to determinism would on its own be sufficient to enable one to get away with locality despite the violation of a Bell-type inequality in the EPR experiment: the commitment to determinism would then imply the rejection of the independence premise, so that there would be no need to abandon also the locality premise when it comes to accounting for the outcomes of the EPR experiment (see Bell 2004, ch. 12). In particular, there would be no point in pursuing a theory such as Bohmian mechanics, whose fundamental law (i.e. the guiding equation) is both non-local and deterministic: the fundamental law should then be either deterministic and local (rejection of the independence premise and acceptance of the locality premise of Bell's theorem) or non-local and indeterministic (rejection of the locality premise and acceptance of the independence premise).

---

[1] I'm grateful to Nicolas Gisin and Travis Norsen for discussions of the topic of this paper and to an anonymous referee for very helpful comments on the original submission.

The result of this investigation will be that the alleged implication from the independence premise to indeterminism does not hold. Nonetheless, there is a subtle issue about the independence premise, which, however, does not concern the issue of determinism vs. indeterminism, but the relationship between generic entanglement and the independent temporal development of subsystems of the universe. In the next section, I consider the two premises of Bell's theorem. Section 3 then examines Bohmian mechanics, comparing this theory to Einsteinian as well as Newtonian determinism. Section 4 goes into an indeterministic quantum dynamics as in the GRW theory.

## 2. *The two premises of Bell's theorem*

Bell's theorem (1964) (reprinted in Bell 2004, ch. 2; see also notably chs. 7 and 24) proves that any theory that complies with the experimentally confirmed predictions of quantum mechanics has to violate either a principle of locality (or local causality) or has to reject the principle that the choice of the variables to be measured on a quantum system in an entangled state can be made independently of what influences the past state of the system (independence principle). The idea behind the locality principle is that, in Bell's words, "The direct causes (and effects) of events are near by, and even the indirect causes (and effects) are no further away than permitted by the velocity of light" (Bell 2004, p. 239). This is one way of formulating the principle of local action that is implemented in classical field theories and that overcomes Newtonian action at a distance.

Consider the simplest version of an EPR-type experiment: two elementary quantum systems are prepared in an entangled state at the source of the experiment (such as two systems of spin 1/2 in the singlet state). Later, when they are far apart in space so that there is no interaction any more between them, Alice chooses the variable to measure in her wing of the experiment and obtains an outcome, and Bob does the same in his wing of the experiment. Alice's setting of her apparatus is separated by a spacelike interval from Bob's setting of his apparatus. The following figure illustrates this situation:

*Figure 1: The situation that Bell considers in the proof of his theorem. Figure taken from Seevinck (2010, appendix) with permission of the author.*

In this figure, *a* stands for Alice's measurement setting, *A* for Alice's outcome, *b* stands for Bob's measurement setting, *B* for Bob's outcome, and λ ranges over whatever in the past may influence the behaviour of the measured quantum systems according to the theory under consideration (which may be standard quantum mechanics, or a theory that admits additional, so-called hidden variables).

The *principle of locality* can then be formulated in the following manner:

$$P(A \mid a, b, B, \lambda) = P_a(A \mid a, \lambda)$$

$$P(B \mid a, b, A, \lambda) = P_b(B \mid b, \lambda) \tag{1}$$

That is to say: the probabilities for Alice's outcome depend only on her measurement setting and λ. Adding Bob's setting and outcome does not change the probabilities for Alice's outcome. The same goes for Bob (see Norsen 2009, 2011 and Seevinck and Uffink 2011 for further precisions concerning the locality principle; and see Goldstein et al. 2011 for an excellent presentation of Bell's theorem).

Furthermore, Bell assumes that the measurement settings *a* and *b* are independent of λ. Let us call this assumption the *independence principle* (2). Thus, *a* and *b* are free variables with respect to λ. Fixing λ in no way restricts the choice of *a* and *b*. In other words, different choices of *a* and *b* can go with the same λ (see Bell 2004, ch. 12, formula (1)). Failure of such independence can arise in two different ways: either the measurement settings exert an influence on λ, or λ somehow influences the measurement settings (or some third factor in the common past correlates λ with the measurement settings). It is obvious from Figure 1 that the first option involves influences travelling backwards in time (see Price 1996, ch. 8 and 9, as well as the papers in *Studies in History and Philosophy of Modern Physics* 38 (2008), pp. 705-784). The second option implies that λ directly (or via a common cause in the common past) constrains the measurement settings, so that these cannot be freely chosen given λ.

The presupposition of an independence between the measurement settings and the prior state of the measured system is not specific for Bell's theorem, but applies to any experimental evidence. To obtain any experimental evidence whatsoever, one has to presuppose that the questions that the experimenter asks (i.e. the choice of measurement settings) are independent of the past state of the measured system. That is the reason why Bell's theorem is widely taken to establish that there is some sort of non-locality in nature, since the locality principle (1) is the only premise that is specific for its derivation (see Maudlin 2011, chs. 1-6, for a detailed assessment leading to this conclusion).

However, this situation would change if it could be established that *any* physical theory whose dynamics is deterministic implies the failure of the independence principle (2). After all, apart from a quantum theory that includes what is known as wave-function collapse in its basic laws, all established physical theories are deterministic. It goes without saying that one can rescue the locality principle (1) despite the results of the EPR experiment by means of determinism: as mentioned above, if one admits influences travelling backwards in time or if one allows λ to have an influence on (or to determine) the choice of the measurement settings *a* and *b*, then one does not have to call the locality principle (1) into question. Nonetheless, backwards causation is a very controversial assumption, and an influence of λ on the measurement settings *a* and *b* is widely considered to involve some sort of conspiracy, making it impossible to examine physical systems in a way that is not predetermined by the past states of these systems. This conspiracy is also known as super-determinism, since it entails that any event in the universe (such as choosing the variables to be measured on a given system) is correlated with any other past event (such as the past state of the system in question) (see Hofer-Szabó, Rédei and Szabó 2013, ch. 9, for a detailed investigation of common cause explanations of the EPR correlations). However, if it could be established that determinism *tout court* implies a violation of the independence principle (2), the objections against rescuing locality by abandoning that principle would lose their force. The situation would then be this one: one could either maintain determinism and would in that case have no reason to abandon the locality principle (1), or one could hold on to the independence principle (2), but would then have to reject *any* deterministic physical theory, including classical mechanics.

*3.    Deterministic dynamics: the case of Bohmian mechanics*

Consider the de Broglie-Bohm theory, going back to de Broglie (1928) and Bohm (1952), whose predominant contemporary version is known as Bohmian mechanics (see the papers in Dürr, Goldstein and Zanghì 2013). This is a deterministic quantum theory of particles moving in three-dimensional space. The theory needs two laws: the guiding equation describing the motion of the particles by means of the wave-function, and the Schrödinger equation describing the temporal development of the wave-function:

In these equations, *Q* denotes the spatial configuration of *all* the particles in three-dimensional space at a given time: $Q(t) = Q_1(t), ... , Q_N(t)$. Hence, $\Psi_t$ is the *universal* wave-function applying to the whole particle configuration of the universe. In the guiding equation, the role of the wave-function, developing according to the Schrödinger equation, is to determine the velocity of each particle at any time *t* given the position of all the particles at that *t*. The theory thereby violates the locality principle (1): to solve the guiding equation, one would have to put in the whole particle configuration of the universe at *t*. Consequently, the guiding equation admits that the motion of any particle at *t* depends, via the universal wave-function, on where all the other particles are located in space at that very *t*.

Because of its dynamics being deterministic, Bohmian mechanics has been taken to violate also the independence principle (2), or a similar principle stipulating that the measurement settings are free variables with respect to the past state of the measured system (see most recently Colbeck and Renner 2011). Conway and Kochen (2009), in their strong free will theorem, include both the locality and the independence principle in their premise MIN, which is rejected in Bohmian mechanics. Bohmians, however, maintain that they abandon only the locality principle in MIN, but not the independence principle (see Goldstein et al. 2010; see also Wüthrich 2011, section 3.2).

Let us first consider what Bell himself says about this situation. He discusses the issue of determinism and the independence principle in the paper "Free variables and local causality" (1977, reprinted in Bell 2004, ch. 12), which is a reply to an objection put forward by Shimony, Horne and Clauser (the full exchange between Bell and them is published in Bell et al. 1985). Bell says:

> Consider the extreme case of a 'random' generator which is in fact perfectly deterministic in nature – and, for simplicity, perfectly isolated. In such a device the complete final state perfectly determines the complete initial state – nothing is forgotten. And yet for many purposes, such a device is precisely a 'forgetting machine'. A particular output is the result of combining so many factors, of such a lengthy and complicated dynamical chain, that it is quite extraordinarily sensitive to minute variations of any one of many initial conditions. It is the familiar paradox of classical statistical mechanics that such exquisite sensitivity to initial conditions is practically equivalent to complete forgetfulness of them. To illustrate the point, suppose that the choice between two possible outputs, corresponding to *a* and *a'*, depended on the oddness or evenness of the digit in the millionth decimal place of some input variable. Then fixing *a* or *a'* indeed fixes something about the input – i.e., whether the millionth digit is odd or even. But this

> peculiar piece of information is unlikely to be the vital piece for any distinctively different purpose, i.e., it is otherwise rather useless. With a physical shuffling machine, we are unable to perform the analysis to the point of saying just what peculiar feature of the input is remembered in the output. But we can quite reasonably assume that it is not relevant for other purposes. In this sense the output of such a device is indeed a sufficiently free variable for the purpose at hand. (Bell 2004, pp. 102-103)

Bell's point can be put in this way: although it is determined by the millionth digit in the initial state of the random generator being odd or even which one of two possible choices of the measurement setting in an EPR experiment is realized, it would amount to conspiracy if λ were correlated with that particular digit. Note that the point that Bell makes in this quotation is about initial conditions only. There is nothing that is specific for the dynamics of quantum mechanics here. Any particular choice of the measurement settings by a random generator corresponds to particular initial positions of the particles that make up the generator. It could happen that these initial particle positions are correlated with the initial positions of the measured particles in the initial configuration of the universe such that, say, whenever the millionth digit in the former is odd, the millionth digit in the latter is even (and *vice versa*). Consequently, there would be correlated measurement outcomes whose correlations are due to particular correlations that happen to obtain in the initial particle configuration of the universe. Such correlations could with reason be regarded as conspiratorial: they require a very specific initial particle configuration of the universe.

However, since this failure of the independence principle (2) obtains in virtue of very specific initial conditions of the universe and not in virtue of the dynamics, it by no means establishes that determinism implies a violation of the independence principle (2). In this case, the correlation between λ and the choice of the measurement settings does not call for a dynamical explanation at all, but is simply due to a very specific initial particle configuration of the universe. In short, it is obvious that such a very specific initial particle configuration is possible, but it is also obvious that nothing follows from this possibility that would link determinism in a systematic way to a violation of the independence principle (2).

Consequently, in order to assess whether such a systematic link obtains, we have to focus on the dynamics by contrast to a specific initial particle configuration. The fact that Bohmian mechanics is deterministic implies that given the particle configuration of the universe at any time and the universal wave-function at that time, the particle configuration at any other past or future time is fixed as well. Furthermore, is clear that, whatever the initial particle configuration of the universe may be, one can conceive an initial wave-function which is such that, when plugged together with the initial particle configuration into the guiding equation and the Schrödinger equation, it will correlate the motion of the particles making up λ with the motion of the particles that later compose the measurement settings from the velocity assigned to the particles in the initial state of the universe on to the time of the performance of the EPR experiment in question. In other words, it is clear that a quantum theory with a deterministic dynamics like Bohmian mechanics admits wave-functions that comply with what is known as super-determinism. In a nutshell, a Bohmian universe can be conspiratorial not only with respect to the initial particle configuration as mentioned above, but also with respect to the dynamics, correlating the past motion of any measured system with the motions of the particles that determine the measurement settings in question so that the independence principle (2) is systematically violated. The issue is whether such an all-comprehensive holism ("everything depends on everything else") is what is to be expected in a Bohmian universe: Is a Bohmian universe such that if you change the position of any arbitrarily picked out particle, you thereby change the velocity of *all* the other particles?

Consider an Einstein universe. Let this be, in our context, a universe that is deterministic, but satisfying the locality principle (1), such as a universe of classical field theory without Newtonian gravitational action at a distance. In such a universe, due to the local dynamics, it would be quite exceptional if the past state of the measured system were correlated with the choice of the measurement settings. This is of course not impossible, since the past light cones of $\lambda$ and the measurement settings overlap (cf. figure 1). However, the mere fact of past light cones overlapping is obviously not sufficient to establish that any future, space-like separated events are correlated. This depends on the dynamics (leaving aside the above mentioned possibility of a conspiratorial initial particle configuration). If the dynamics is local, there are many initial states that are identical as far as $\lambda$ is concerned and that differ in the initial states of the particles that later compose the measurement settings (or the random generator or the brain of the experimental scientist), resulting therefore in different settings of the measurement apparatuses. In other words, in an Einstein universe, it is no problem to hold $\lambda$ fix and to change only the initial positions and velocities of the particles that later compose the measurement settings, and different measurement settings will be the consequence. In a nutshell, no correlation between $\lambda$ and the measurement settings, because same $\lambda$ and different measurement settings are possible: same $\lambda$ by no means implies same measurement settings, and a difference in the measurement settings by no means implies a difference in $\lambda$.

We can therefore put down the following first result of this paper: *It is not determinism as such that poses a challenge to the independence principle in the derivation of Bell's theorem.* A local deterministic dynamics does not do so. By way of consequence, *it is only through the combination of determinism and non-locality that the issue of such a challenge can arise*. Hence, one cannot get away with locality by simply rejecting the independence principle (2). If one abandons that principle in a local and deterministic universe, one has to make a very specific assumption about the initial dynamical state of the universe so that one obtains not only a deterministic, but also a super-deterministic universe. One thereby falls back into the specific problems that super-determinism poses ("conspiracy"), but that do not touch determinism as such in a local deterministic universe.

Let us now briefly turn to Newtonian mechanics with gravitational action at a distance. A Newtonian universe is deterministic and non-local, although Newtonian non-locality is different from quantum non-locality: in a Newtonian universe, the distribution of mass in any given region of space instantaneously affects the motion of the particles in any other region of space, but that effect goes down with the square of the distance, so that in most cases, it becomes negligible as the distance increases. Nonetheless, the initial distribution of mass (that is, the initial dynamical state) in a Newtonian universe can be such that the masses in $\lambda$ influence the motion of the masses that will make up the measurement settings (or the random generator or the brain of the experimental scientist), so that the distribution of the masses in $\lambda$ determines a specific setting in the EPR experiment. However, again, in order for this result to obtain, one has to presuppose a very specific initial dynamical state of the universe, namely a very specific distribution of the masses. In general, there has to be no variation in $\lambda$ for there to be a difference in the measurement settings: in a Newtonian like in an Einsteinian universe, it is no problem to hold $\lambda$ fix and to change only the initial positions and velocities of the particles that later compose the measurement settings (or the random generator or the brain of the experimental scientist), and different measurement settings will be the consequence.

A central difference between quantum (including Bohmian) non-locality and Newtonian non-locality is that the former does not systematically depend on the distance between the particles. In the EPR experiment, the measurement outcomes in the two wings are correlated whatever the spatial distance between the two wings is. Furthermore, entangled states are generic in Bohmian mechanics, as they are in quantum mechanics in general. One may therefore be inclined to develop the following reasoning: since entanglement is generic, one has to assume that the state of the initial particle configuration of the universe is an entangled state in such a way that the velocity of *any* particle in the initial state of the universe depends on the position of *all* the other particles in the universe via the initial universal wave-function. Furthermore, since the initial state of the universe fixes all the future temporal development, the future positions of any particle are thereby correlated with the future positions of any other particle. Only this conclusion yields the result that also the positions of the particles that make up the measurement settings (or the random generator or the brain of the experimental scientist) are correlated with the positions of the measured particles: there is nothing in particular that these two systems of particles have in common that would distinguish them from all the other particles in the universe (unless one considers a conspiratorial initial particle configuration as mentioned above). Consequently, for them to be correlated, the motions of all the particles in the universe have to be correlated with one another.

If this reasoning is sound, then the independence principle (2) is violated in a Bohmian universe. Note, however, that this violation, if indeed it obtains, is a consequence of the particular form that quantum non-locality takes (entangled states). Hence, it could not be employed to circumvent that non-locality. Nonetheless, this reasoning is not conclusive. It takes generic entanglement as far as the universal wave-function is concerned to be a sufficient condition for a violation of the independence principle (2), but that implication does not hold. The reason is that Bohmian mechanics endorses not only a state of the universe that is an entangled state and a universal wave-function that never collapses, but on this very basis allows at any time for the conception of effective wave-functions for subsystems of the universe and effective wave-function collapse (see Dürr, Goldstein and Zanghì 2013, in particular chs. 2 and 5, for the details how this is done). Effective wave-functions have not only a pragmatic status for all practical purposes, but also an ontological one: if they apply to subsystems of the universe, they show that the temporal development of these subsystems (their motion) occurs independently of what there is in the rest of the universe, as long as effective wave-functions apply to them. In other words, in Bohmian mechanics, generic entanglement on the level of the universal wave-function goes together with effective wave-functions for subsystems of the universe, signifying that these subsystems develop in time independently of each other, although it is never in principle excluded that their motions can at some time become correlated.

Thus, prior to the measurement in an EPR-type experiment, also in a Bohmian universe, it is a well-grounded assumption that the temporal development of the particles making up the measurement apparatuses (or the random generator or the brain of the experimental scientist) and the particles to be measured can be described by effective wave-functions each. That is to say: the particles to be measured (or the particles in λ in general) and the particles making up the measurement apparatuses develop independently of each other. Hence, also in a Bohmian universe, a violation of the independence principle (2) implies a very specific dynamical initial state of the universe, namely an initial state of the universe that rules out effective wave-functions applying to the particles that will later compose the measurement settings (or the random generator or the brain of the experimental scientist choosing the measurement settings) and the particles that will later be measured (again leaving aside the possibility of a conspiratorial initial particle configuration as mentioned above).

Consequently, also in Bohmian mechanics, there is a crucial difference between the following two situations: (a) the two particles in the EPR-experiment are entangled, so that there are no effective wave-functions describing the temporal development of each of these particles independently of the other one; (b) that particle pair and the measurement settings (or the random generator or the brain of the experimental scientist choosing the measurement settings) are described by effective wave-functions each, so that they develop independently of each other prior to the measurement. Only if that latter independence were violated so that there were no effective wave-functions, Bohmian mechanics would violate the independence principle (2).

The second result of this paper is hence this one: *also in a Bohmian universe, quantum non-locality does not imply that the independence principle (2) is violated; generic entanglement as far as the universal wave-function is concerned notwithstanding, there are effective wave-functions for subsystems of the universe, signifying that the temporal development of these subsystems occurs independently of each other*. In brief, and of course setting aside the possibility of conspiratorial initial conditions, even if non-locality is combined with determinism, the independence principle (2) is not called into question, as long as there are effective wave-functions for subsystems of the universe.

## 4.    Indeterministic dynamics: the case of GRW

Let us now briefly go into an indeterministic quantum dynamics. Consider the GRW quantum theory, going back to Ghirardi, Rimini and Weber (1986). In brief, the GRW theory modifies the Schrödinger equation in a non-linear and stochastic way such that this equation includes wave-function collapse. There are two proposals for a quantum ontology of matter in three-dimensional space in the framework of this theory. The one proposal, put forward by Ghirardi himself, postulates a continuous matter density field that develops according to a GRW-type equation with a continuous spontaneous localization of the wave-function, describing the spontaneous concentration of the matter density field in certain regions of space (see Ghirardi, Grassi and Benatti 1995; and see Ghirardi, Pearle and Rimini 1990 as well as Gisin 1989 for this sort of dynamics). The other proposal, going back to Bell (2004, ch. 22, originally published 1987) and later elaborated on by Tumulka (2006, 2009), assumes that there are only single events at certain points of physical space, called "flashes", and works with a GRW-type equation with jumps of the wave-function: in brief, the spontaneous localization of the wave-function in configuration space describes the occurrence of flashes in physical space. The differences between these proposals as well as between various GRW-type equations are irrelevant for present purposes. The decisive point is that any GRW universe is indeterministic.

However, the GRW indeterminism is not to be confused with randomness. In a GRW universe, given the initial configuration of the matter density field or the flashes and the initial wave-function, possible entire histories of the development of the matter density field or the flashes are fixed by the initial wave-function figuring in the GRW-equation, and probabilities are assigned to these histories. That is the reason why the GRW probabilities are objective; applied to the universal wave-function, they concern possible histories of the universe given an initial configuration of matter and an initial wave-function. Consequently, one can with good reason maintain that the GRW probabilities refer to propensities for a certain development of the universe that are inherent in the initial state of the universe, making certain developments of the universe more probable than others (cf. Dorato and Esfeld 2010; see also Gisin 1991).

The crucial difference between the GRW theory and Bohmian mechanics then is that in a GRW universe, the universal wave-function collapses and thereby erases the traces of all past correlations: a GRW universe is time-asymmetric. By contrast, in Bohmian mechanics, given the particle configuration of the universe and the universal wave-function at any time, the past motion of the particles is fixed in the same way as their future motion. Hence, in a GRW universe, whatever correlations there are in the past state of the universe, traces of these correlations will in general be erased by future wave-function collapse. However, this does not make a significant difference as far as our case is concerned: GRW collapse takes the place of effective wave-functions in Bohmian mechanics (and *vice versa*). Both lead to the same result in our context, namely that quantum entanglement does not imply a violation of the independence principle (2), since both imply that subsystems of the universe develop in time independently of one another. The third result of this paper therefore is this one: *the issue of determinism vs. indeterminism is irrelevant to the issue of whether or not quantum non-locality (violation of the locality principle (1)) implies a failure of the independence principle (2)*.

5.   *Conclusion*

This paper has sought to establish three results:

1. It is not determinism as such that poses a challenge to the independence principle in the derivation of Bell's theorem. It is only through the combination of determinism and non-locality that the issue of such a challenge arises.
2. Also in a Bohmian universe, quantum non-locality combined with determinism does not imply that the independence principle (2) is violated; generic entanglement as far as the universal wave-function is concerned notwithstanding, there are effective wave-functions for subsystems of the universe, signifying that the temporal development of these subsystems occurs independently of each other.
3. The issue of determinism vs. indeterminism is irrelevant to the issue of whether or not quantum non-locality implies a failure of the independence premise.

Hence, Bell's theorem tells us nothing about determinism or indeterminism. *A fortiori*, it has no bearing on free will. As regards free will, the majority view in contemporary philosophy is compatibilism, that is, the stance that there can be free will even in a deterministic universe, since free will is distinct from randomness: the decisive issue is which parameters (or reasons) form the will of a person, such that they are compatible with the autonomy of the person, and not the absence of any such parameters (cf. e.g. Frankfurt 1971). By contrast, if one rejects compatibilism, going for what is known as libertarianism, one is in trouble with any laws of physics, be they deterministic or indeterministic, since even the latter fix objective probabilities for the development of the entire universe including the physical movements of persons (as in the GRW theory), which are independent of the minds of persons or any mental variables in general.

Acknowledgements: I'm grateful to Nicolas Gisin and Travis Norsen for discussions of the topic of this paper and to an anonymous referee for very helpful comments on the original submission.
Conflict of Interest: The author declares that he has no conflict of interest.